\documentclass[11pt,twoside]{article}
\usepackage{asp2010}

\resetcounters

\newcommand{\h}{{\it Herschel}}

\newcommand{\lsol}{L$_{\odot}$}
\newcommand{\lir}{$L_{\rm IR}$}

\newcommand{\mips}{$S_{\rm 24}$}

\newcommand{\td}{$T_{\rm d}$}

\begin{document}

\title{Towards a complete census of high-z ULIRGs with \h}
\author{Georgios E. Magdis$^1$, D. Elbaz$^1$, H.S. Hwang$^1$,PEP \& HerMES team
\affil{$^1$CEA Saclay}
%\affil{$^2$Institution Full Address for Author2}
%\affil{$^3$Institution Full Address for Author3}}\ref{\pageref{\index{\cite{\footnote{•}}}}}
}
\begin{abstract}
Using Herschel PACS and SPIRE observations as part of the HerMES project, we explore the far-IR properties of a sample of mid-IR selected starburst dominated ultra-luminous infrared galaxies (ULIRGs) at z $\sim$ 2. We derive robust estimates of infrared luminosities (\lir) and dust temperatures (\td) of the population and find that galaxies in our sample range from those that are as cold as high-z sub-millimeter galaxies (SMGs) to those that are as warm as optically faint radio galaxies (OFRGs) and local ULIRGs. We also demonstrate that a significant fraction of our sample would be missed from ground based (sub)mm surveys (850-1200$\mu$m) showing that the latter introduce a bias towards the detection of colder sources. Similarly, based on PACS data as part of the PEP project, we construct for the first time the full average SED of a sub-sample of infrared luminous Lyman break galaxies at z$\sim$3, and find them to have higher $T_{\rm d}$ when compared to that of SMGs with comparable \lir. We conclude that high-z ULIRGs span a wide range of dust temperatures, larger than that seen in local ULIRGs, and that Herschel data provide the means to characterize the bulk of the ULIRG population, free from selection biases introduced by ground based (sub)mm surveys. 
\end{abstract}

\section{Introduction}
A key parameter to study galaxy evolution is to probe the 
census of the star formation activity, both in the distant and local universe.
To this direction, it has been shown that the contribution of luminous infrared 
galaxies (\lir ~$>$ 10$^{11}$ \lsol) to the star formation density 
is progressively rising as we look back in the comic time, at least up to $z$ $\sim$ 2. Indeed, 
although they were found to be rare in the local universe and to account only for $\sim$ 5$\%$ of 
the total infrared energy emitted by galaxies at $z$ $\sim$ 0, LIRGs along with the ULIRGs (\lir ~$>$ 10$^{12}$ \lsol), dominate the SFR density at $z$ $\sim$ 1-2, accounting for the 70$\%$ of the star formation activity at these epochs (e.g. Le Floc'h et al. 2005).

Until recently, the most successful methods for selecting high-z ULIRGs was their direct far-IR detection via ground based
(sub)millimeter surveys (e,g, Hughes et 1996). However, the submillimetre technique introduces a bias towards the selection
of ULIRGs with lower dust temperatures while it misses
warmer ULIRGs. First observational evidence of a missing
population of high-z dusty star-forming galaxies with
hotter dust  was been given by Chapman et al. (2004) 
using a selection of radio-detected but sub-mm-faint galaxies (OFRGs) with UV spectra consistent with high-z starbursts. Recent studies (Magnelli et al. 2010 and Chapman et al. 2010) have shown that their is no overlap between the two populations (SMGs and OFRGs) in the \lir ~ \td ~space, leaving a large gap between them. In other words, it appears that if SMGs are biased towards the colder high-z ULIRGs, then OFRGs trace only the ULIRGs with warmest \td .
These suggest that in order to obtain the complete census of high-z ULIRGs, a selection independent of the sub-mm emission should  be employed.

Another technique that has been proven to pick high-z
starburst dominated ULIRGs efficiently, is based on mid-IR
color selection, fine tuned to probe the rest-frame 1.6$\mu$m bump, prominent in the SED of star-forming galaxies. 
Among others (e.g. Farrah et al. 2003), Huang et al. (2009) demonstrated that a simple  IRAC color criteria: $0.05<[3.6]-[4.5]<0.4$ and $−0.7<[3.6]-[8.0]<0.5$,  coupled with a \mips ~flux cut (\mips ~$>$0.2mJy), selects 
star-burst dominated ULIRGs in a narrow redshift range of 1.5 $<$ $z$ $<$ 2.5. Using Herschel PACS and SPIRE data, as part of the HerMES (Oliver et al. 2011 in prep) project we 
explore the far-IR properties of the population, derive robust \td ~and \lir ~measurements for the bulk
of the population and compare our sample to that of other high-z ULIRGs. All results, figures and discussions here, are presented in detail in Magdis et al. (2010a,b).
\begin{figure*}
\centering
\includegraphics[scale=0.22]{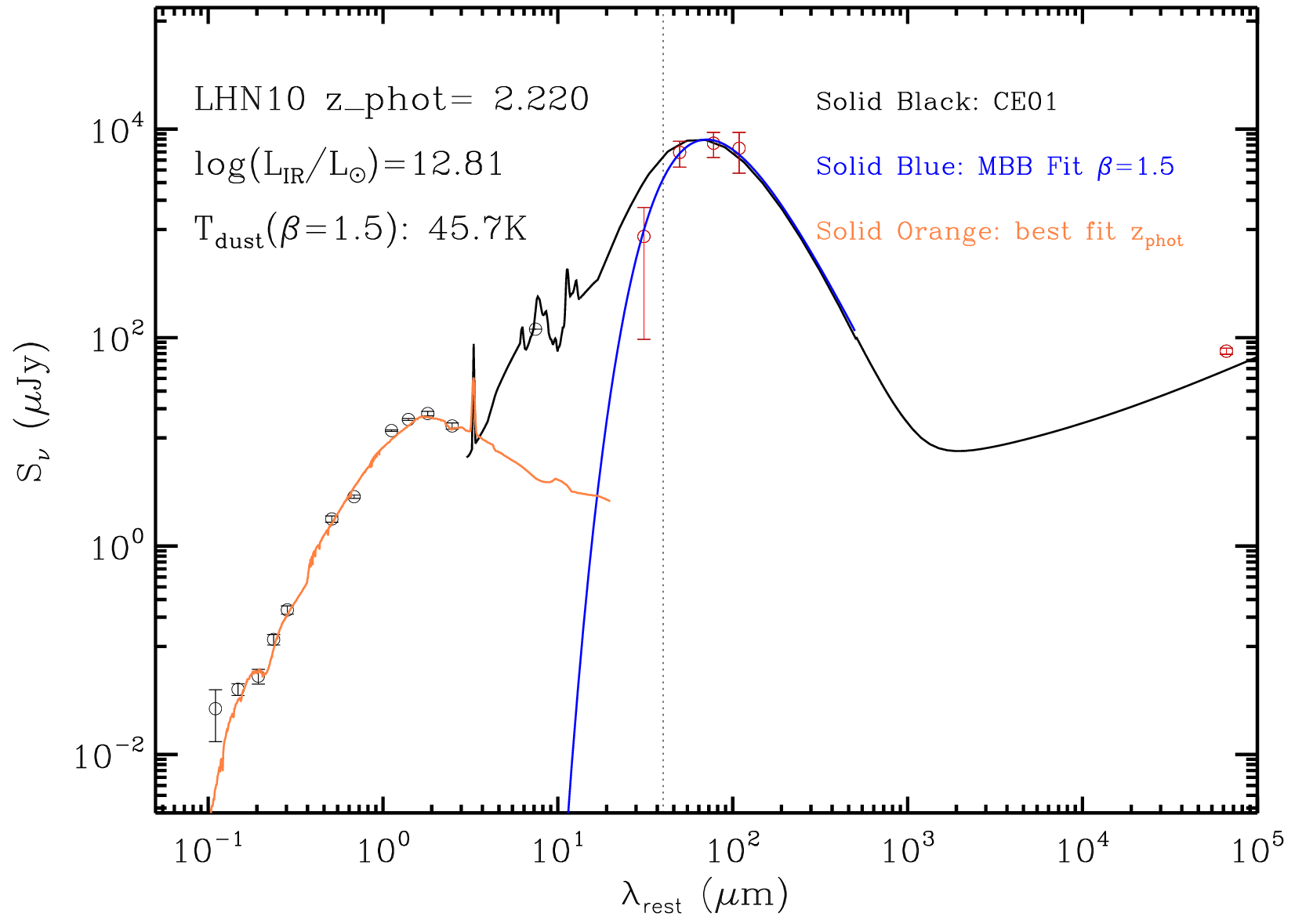}
\includegraphics[scale=0.22]{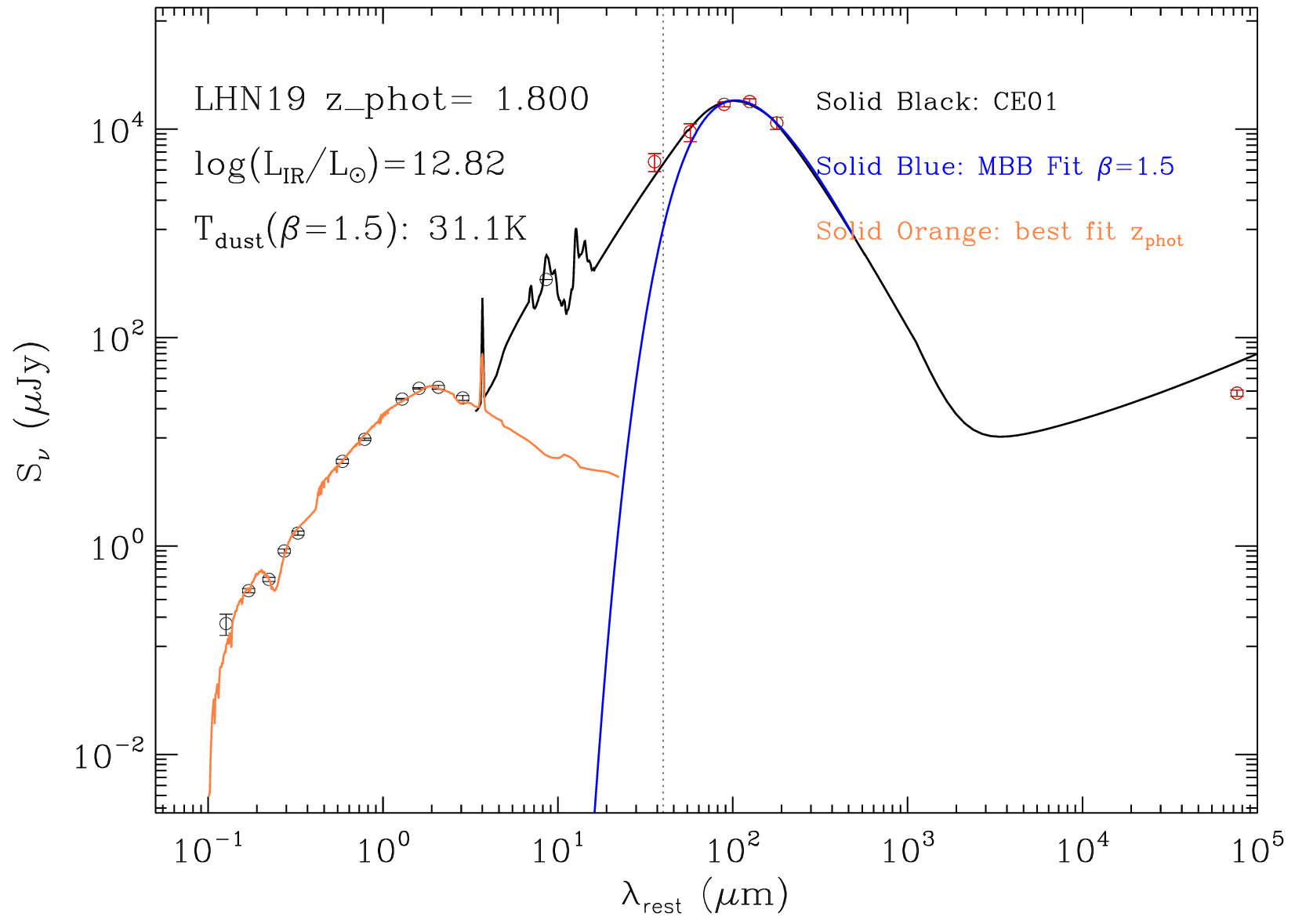}\\
\caption{\small{Rest-frame SEDs and derivation of the far-IR properties for two ULIRGs in our sample. Solid orange line shows the best fit
template up to observed 8$\mu$m as derived by LEPHARE photo-z code. Solid black line shows the best fit CE01 model while the blue line
depicts the best-fit modified black body (with $\beta$=1.5), used to derive \td ~estimates. The vertical dotted line indicates the wavelength cut,
below which photometric data where not considered in the modified black body fit. Red circles denote it Herschel data}}
\label{fig:sfsed}
\end{figure*}
\section{Far-IR properties of IRAC selected ULIRGs}
We select sources with $0.05<[3.6]-[4.5]<0.4$, $-0.7<[3.6]-[8.0]<0.5$,  and \mips ~$>$ 0.2mJy in Lockman Hole and GOODS-N, 
and with at least two detections in the either of PACS and SPIRE bands. Our sample consist of 25 ULIRGs with a median $z=2.01$
and with 18 out of 25 objects lying in narrow redshift range (1.7 $<$ $z$ $<$ 2.3) (for details see Magdis et al. 2010a). To derive estimates for the \lir ~of the galaxies in
our sample, we first convert their SED to rest-frame applying k-corrections and then fit the PACS and SPIRE data with
the libraries of Chary \& Elbaz (2001) (CE01) and Dale \&
Helou (2002). Results based on the two methods are in very close agreement indicating a median \lir ~= 3 $\times$ 10$^{12}$ \lsol. 
To derive the dust temperature of galaxies in our sample, we use a single temperature modified black body fitting form. This model was fit to \h ~data with rest-frame $>$ 40$\mu$m, assuming a fixed emissivity index of $\beta$=1.5. Two examples of the rest–frame SEDs along with the best-fit CE01 templates for two ULIRGs in our sample are shown in Figure 1. We find that our sample spans in wide range of dust temperatures 25$<$ \td $<$ 62 (K), while the luminosities vary by less than an order of magnitude 12.24 $<$ log(\lir / \lsol) $<$ 12.94. The median values are \td ~= 42.3 K, and \lir ~=3 $\times$ 10$^{12}$ \lsol .

\section{Herschel reveals a \td ~unbiased selection of $z$ $\sim$ 2 ULIRGs}
We now compare the far-IR properties of our sample with that of local and high-$z$ ULIRGs selected by different techniques 
We consider the large set of $z$ $\sim$ 2 SMGs (Chapman et al. 2005 and Kovacs et al. 2006), a sample of $z$ $\sim$2 OFRGs
(Casey et al. 2009, Magnelli et al. 2010) and a compilation of local/intermediate-$z$ (0$<$ $z$ $<$ 0.98) ULIRGs (Clements et al. 2010, Farrah et al.
2003 and Yang et al. 2007). In all these studies, the method
to derive \td ~estimates is similar to ours, fitting modified
black-body models to the far-IR photometric points and assuming $\beta$ = 1.5. This comparison is illustrated in Figure 2(top).

Our observations confirm the existence of ULIRGs in the high-z universe with dust temperature higher than that of SMGs. 
Furthermore, it seems that the selection of high-z ULIRGs based on the detection of the 1.6$\mu$m bump does not favour a particular \td, selecting ULIRGs that overlap with the SMGs and OFRGs but also ULIRGs of intermediate \td. Indeed, for the luminosity bin of our sample, SMGs have a median \td ~= 36 $\pm$ 8 K while OFRGs are considerably warmer with median \td ~=47$\pm$ 3 K (Magnelli et al 2010) while galaxies in our sample range from those that are as cold as SMGs 
to objects as warm as OFRGs, with a significant fraction located in the intermediate region between the two samples, bridging the two populations. 
We also note that a large fraction of the sample falls in the \td ~- \lir  ~relation 
of the local ULIRGs. Finally, our data indicate that the \td ~dispersion of high-$z$ ULIRGs is larger than that of the local ULIRGs as derived
based on IRAS/AKARI observations. A similar conclusion is reached by Hwang et al. (2010).

We also  estimate the $S_{\rm 850}$ flux densities of our sample based on the best fit CE01 model. The predicted  $S_{\rm 850}$  fluxes of our sample along
with the measured sub-mm flux of high-z SMGs are plotted versus the derived \td ~of the two populations in Figure
2 (bottom). We also overplot tracks in constant \lir. This plot illustrates that a significant fraction (60\%) of the mid-IR selected ULIRGs in our sample have $S_{\rm 850}$ flux densities lower than that of the SMGs, 
lie below the confusion limit at 850$\mu$m (2-3 mJy) and hence would
be missed by ground-based (sub)mm surveys. Nevertheless,
we also find IRAC-peakers with predicted  $S_{\rm 850}$  above the
detection limit and which therefore should be detected in
the sub-mm. Observational data confirm our results, with $\sim$40\% of the sample being detected in the submm bands 
(e.g. Lonsdale et al. 2009). We conclude that our analysis strongly suggests that \h ~data allow us
 for the first time to characterize the far-IR properties of 50\% of the mid-IR selected ULIRGs that would be missed by ground based (sub)mm surveys
and reveal that their properties are different from that of SCUBA/IRAM selected galaxies.
\begin{figure*}
\centering
\includegraphics[scale=0.35]{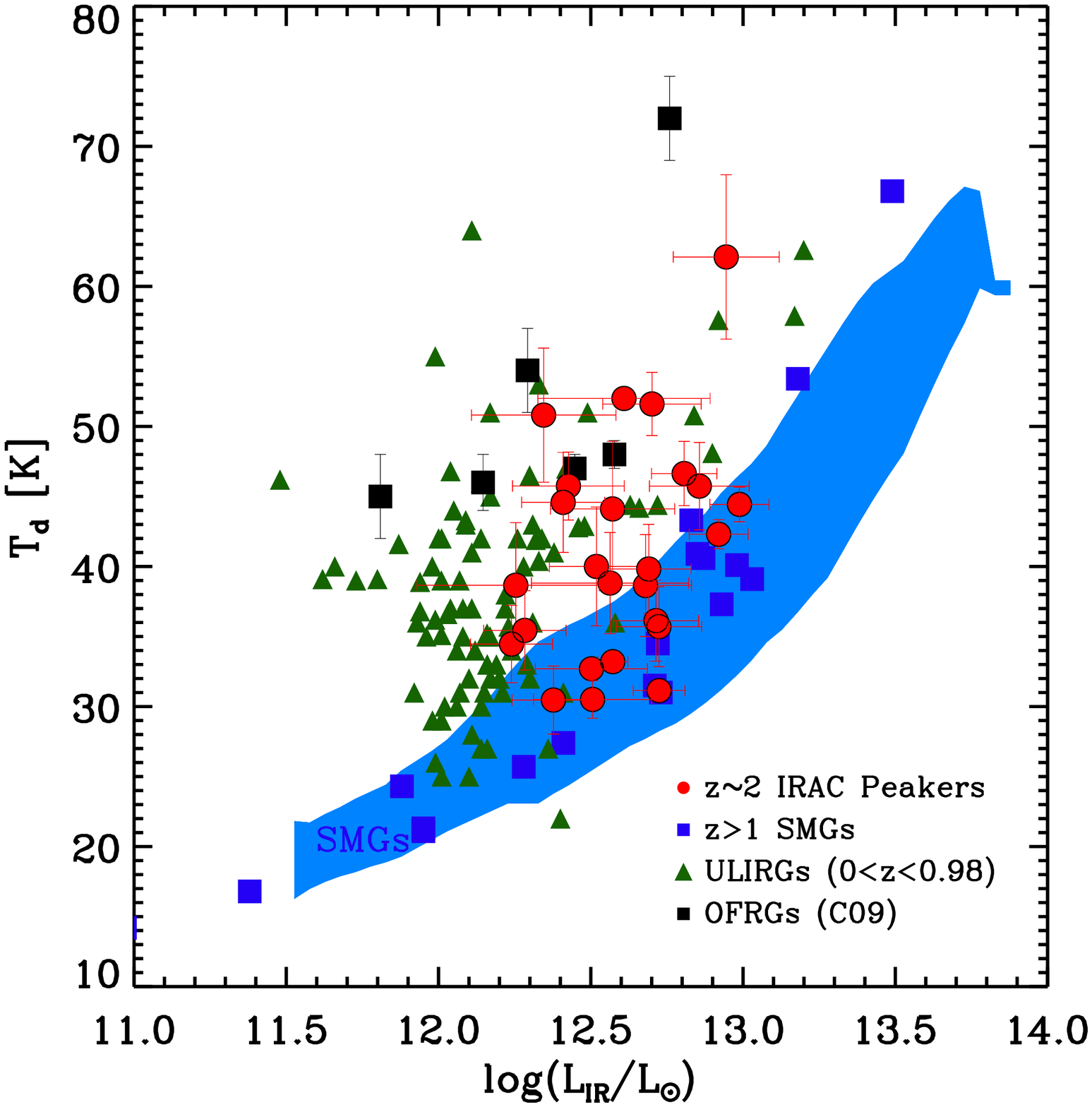}
\includegraphics[scale=0.35]{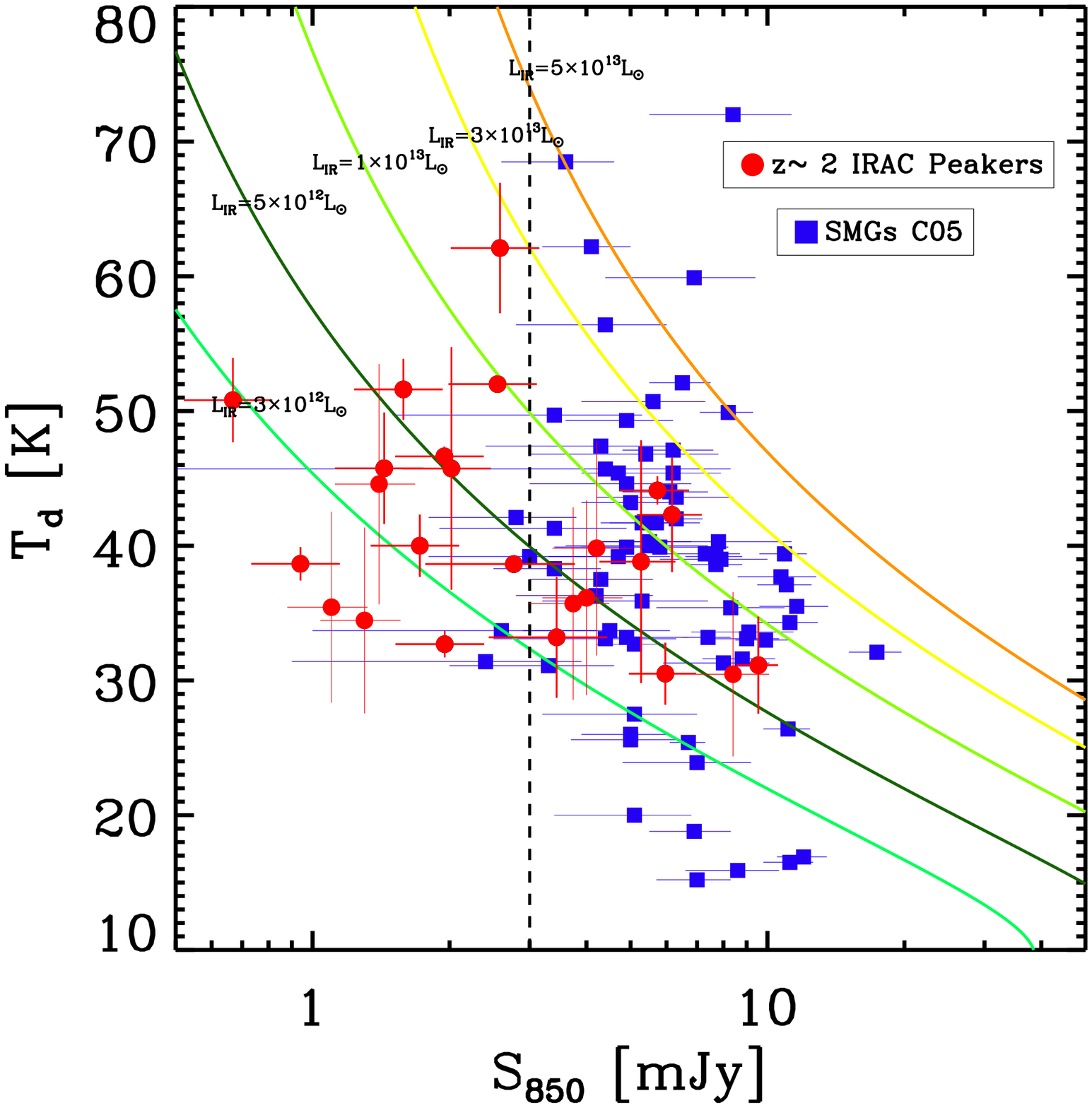}\\
\caption{\small{top) The \lir -  \td ~relation for IRAC selcted ULIRGs (red circles).
Included are results for local/intermediate-$z$ ULIRGs (green filled
triangles, Farrah et al. 2003, Clements et al. 2010, Yang et al.
2007), high-$z$ SMGs (blue squares, Chapman et al. 2005, Kovacs
et al. 2006) and OFRGs (black squares, Casey et al. 2009). The
cyan shaded area denotes the 2$\sigma$envelope of the \lir ~- \td ~relation
of high-$z$ SMGs. For a given \lir, our sample span in a wide
range of dust temperatures, bridging the ``cold'' high-z SMGs to
the ``warmer'' local/intermediate-$z$ ULIRGs and $\sim$2 OFRGs. bottom) \td ~versus the estimated $S_{\rm 850}$ flux densities of galaxies 
in our sample (red circles). We also include \td
~measurements and observed $S_{\rm 850}$ flux densities of high-$z$ SMGs by
Chapman et al, (2005) (blue squares). Solid lines represent tracks
in constant \lir ~while the vertical dotted line indicates the confusion limit of current ground based submm surveys. 
It is evident that a significant fraction of our sample lies below the detection
limit and would be missed the SCUBA-850$\mu$m surveys, if we consider that the detection limit should be above the confusion.}}
\label{fig:sfsed}
\end{figure*}
\begin{figure*}
\centering
\includegraphics[scale=0.35]{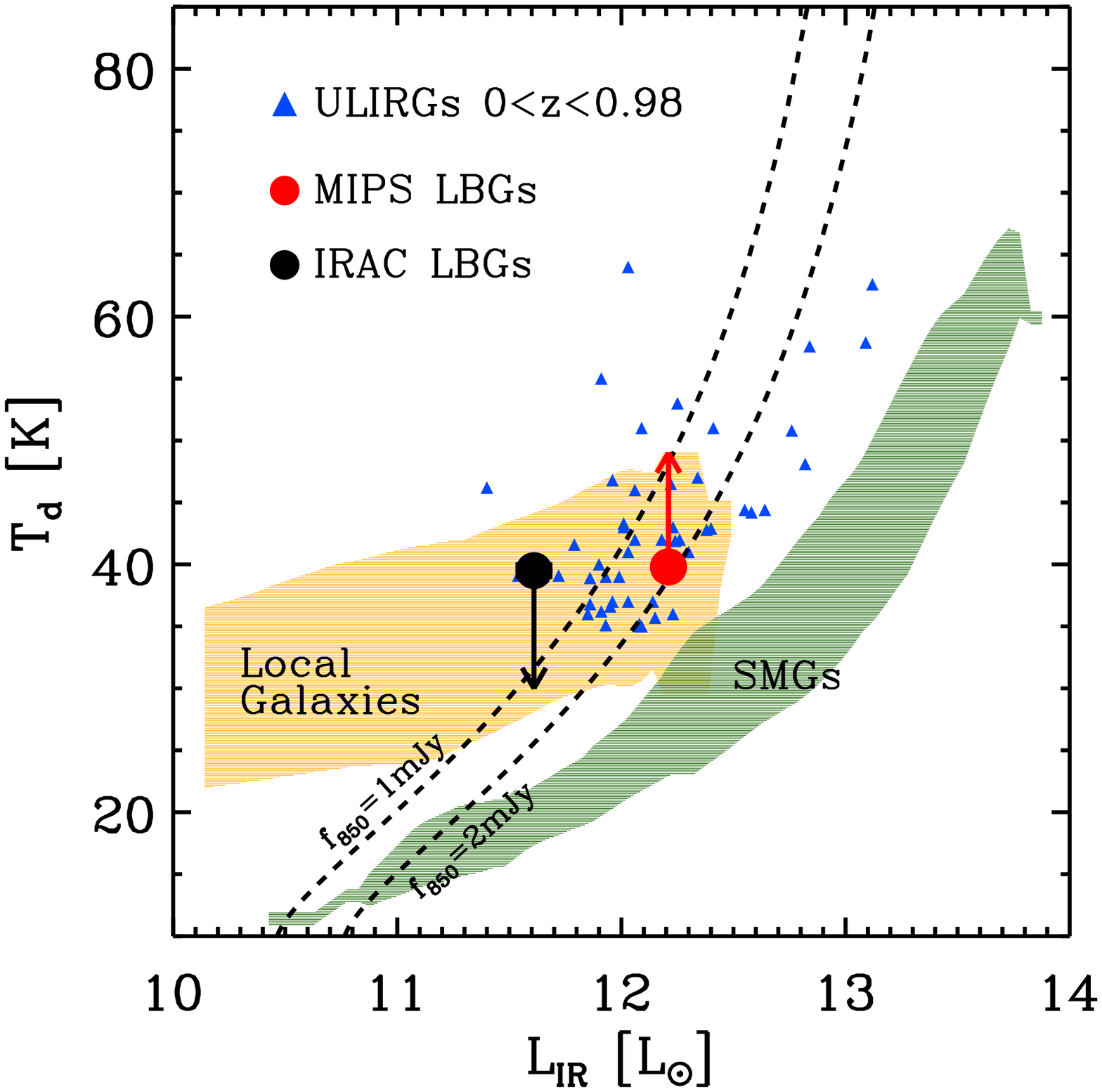}\\
\caption{\small{The \td ~- \lir ~relation for MIPS and IRAC LBGs (red and black circles respectively).
We note that these values correspond to 3 $\sigma$ upper and lower \td ~limits for IRAC and MIPS
LBGs. The green shaded region depicts the loci of high-$z$ SMGs by Chapman
et al. (2005), while the orange shaded area shows the 2$\sigma$ envelope of the \lir ~- \td ~relation
for local IR galaxies in SDSS adopted from Hwang et al. (2010).
Black dashed lines represent tracks of constant flux density at 850$\mu$m ($S_{\rm 850}$=1mJy and $S_{\rm 850}$=2
mJy) for galaxies at z=3. Objects at z=3 with higher $S_{\rm 850}$, lie on the right of the lines. Adopted from Magdis et al. (2010b)}}
\label{fig:sfsed}
\end{figure*}
\section{\td ~of LBGs at z$\sim$3 }
Similar results were reached, when we considered, a sub-sample of 24$\mu$m detected Lyman break galaxies at z$\sim$3 (MIPS LBGs).  
Using PACS data as part of the PEP project, we derived a median IR luminosity of \lir ~= 1.6 $\times$ 10$^{12}$ \lsol , 
placing these galaxies in the class of ULIRGs (Magdis et al. 2010b,c). Considering the large \lir ~and the substantial dust reddening of these LBGs 
it is somewhat surprising that there are only few examples of direct sub-millimeter
detection for these galaxies. MIPS-LBGs are the most rapidly star-forming, most luminous,
and dustiest galaxies among the high redshift UV-selected population, and therefore are
the best candidates for having far-IR emission that could be detected in current sub-mm
surveys. Based on the average SED of MIPS-LBGs as constructed by stacking at 
PACS, Aztec1.1mm  and VLA1.4GHz maps (Magdis et al. 2010b),  
we predict that the flux density of the MIPS-LBGs emitted
at 850$\mu$m is $S_{\rm 850}$ =1.1-1.5mJy, just below the current confusion limit. It could therefore be
suggested that MIPS-LBGs provide a link between SMGs and typical UV selected LBGs
that are faint in the IR.

In Figure 3 we compare dust temperature versus infrared luminosity for the MIPS LBGs with that of the z $\sim$ 2 SMGs 
by Chapman et al. (2005). We also plot, the 3$\sigma$ envelope of the \lir ~- \td ~relation for local infrared galaxies in SDSS (Hwang et al. 2010).
It is evident that for the \lir ~of the MIPS-LBGs, the bulk of SMGs are considerably
colder, while MIPS-LBGs fall in the locus of the local ULIRGs and are within the scatter
observed in local galaxies. Based on modified black body models, we also compute tracks
of constant 850$\mu$m flux density for galaxies at z=3, close to the confusion/detection limit of
current sub-mm surveys ($S_{\rm 850}$=1mJy and $S_{\rm 850}$=2mJy). MIPS-LBGs lie in between the two
tracks, indicating that a typical MIPS detected LBG emits at 1-2 mJy level at the sub-mm
bands. This explains the small overlap between the LBGs and SMGs found
in previous studies.

\section{Conclusions}
Based on \h ~ observation of z$\sim$2 and $\sim$3 ULIRGs, as part of the PEP and HerMES projects, we explore the far-IR properties of these samples and find that :

\begin{itemize}
\item  IRAC selected ULIRGs display a wide range of \td, ranging from those that are as cold as high-$z$ SMGs to objects as warm as OFRGs, 
while a significant fraction has intermediate \td, bridging the two populations. This indicates that the mid-IR selection of high-$z$
ULIRGs does not introduce a systematic bias in \td.
\item A significant fraction of z $\sim$ 2 ULIRGs are missed from (sub)mm surveys, showing that the sub-mm technique introduces a bias towards the detection of
colder ULIRG sources. On the other hand, \h ~ data provide the means for a complete and unbiased selection of the census of ULIRGs at this redshift.
\item The \td ~dispersion of high-z ULIRGs is larger than that found in the local universe, indicating a wide range of mechanisms triggering the star formation activity at earlier epochs.
\item Infrared luminous LBGs at z $\sim$ 3, have warmer \td ~than SMGs galaxies while they fall in the locus of the \lir ~-\td ~relation of the local
ULIRGs. This, along with estimates based on the average SED, explains the marginal detection of LBGs in current sub-mm surveys and suggests that these
latter studies introduce a bias towards the detection of colder ULIRGs in the high-z universe, while missing high-z ULIRGs with warmer dust.
\end{itemize}
\acknowledgements Herschel is an ESA space observatory with science instruments provided
  by European-led Principal Investigator consortia and with important participation from NASA. 
  This study is based on data obtained as part of the HerMES and PEP  \h ~ projects.

\end{document}